# Flame/flow dynamics at the piston surface of an IC engine measured by high-speed PLIF and PTV

C.-P. Ding[1], B. Peterson[2], M. Schmidt[1], A. Dreizler[1], B. Böhm[1]

[1]Fachgebiet Reaktive Strömungen und Messtechnik (RSM), Technische Universität Darmstadt, Otto-Berndt-Str. 3, 64287 Darmstadt, Germany

[2]School of Engineering, Institute for Multiscale Thermofluids, University of Edinburgh, The King's Buildings, Mayfield Road, Edinburgh, EH9 3BF, UK

**Corresponding author:** Name: Brian Peterson, Address (see [2]), Fax: , E-mail: brian.peterson@ed.ac.uk

**Abstract**

Resolving fluid transport at engine surfaces is required to predict transient heat loss, which is becoming increasingly important for the development of high-efficiency internal combustion engines (ICE). The limited number of available investigations have focused on non-reacting flows near engine surfaces, while this work focuses on the near-wall flow field dynamics in response to a propagating flame front. Flow-field and flame distributions were measured simultaneously at kHz repetition rates using particle tracking velocimetry (PTV) and planar laser induced fluorescence (PLIF) of sulfur dioxide ($SO_2$). Measurements were performed near the piston surface of an optically accessible engine operating at 800 rpm with homogeneous, stoichiometric isooctane-air mixtures. High-speed measurements reveal a strong interdependency between near-wall flow and flame development which also influences subsequent combustion. A conditional analysis is performed to analyze flame/flow dynamics at the piston surface for cycles with 'weak' and 'strong' flow velocities parallel to the surface. Faster flame propagation associated with higher velocities before ignition demonstrates a stronger flow acceleration ahead of the flame. Flow acceleration associated with an advancing flame front is a transient feature that strongly influences boundary layer development. The distance from the wall to 75% maximum velocity ($\delta_{75}$) is analyzed to compare boundary layer development between fired and motored datasets. Decreases in $\delta_{75}$ are strongly related to flow acceleration produced by an approaching flame front. Measurements reveal strong deviations of the boundary layer flow between fired and motored datasets, emphasizing the need to consider transient flow behavior when modelling boundary layer physics for reacting flows.

# Flame/flow dynamics at the piston surface of an IC engine measured by high-speed PLIF and PTV

C.-P. Ding, B. Peterson, M. Schmidt, A. Dreizler, B. Böhm

## 1. Introduction

The topic of near-wall reacting flows has recently gained significant attention for the development of modern internal combustion engines (ICE) with lower cylinder volumes and higher power densities. These downsized, boosted ICE concepts provide greater efficiency, but are subject to challenges such as increased transient heat transfer and flame quenching, which limit efficiency gains. A detailed understanding of the physical processes of mass, momentum, and heat transfer (and their mutual interactions) near surfaces is required for successful development of modern ICE technologies.

In spark-ignition engines, gas phase convection is regarded as the primary mechanism of near-wall heat transfer [1-3]. The accuracy with which one can predict this energy transfer often lies within the ability to resolve the velocity boundary layer. ICE simulations based on LES and RANS require wall models to compute wall shear stress and heat flux gradients in the thin boundary layer. Most models are based on the law-of-the-wall, originally derived from steady channel flow [4]. Recent experimental and DNS studies report strong deviations of engine boundary layers compared to the law-of-the-wall [5,6] such that the predictive capabilities of present wall modelling approaches in ICEs are rather limited and investigations of near-wall flows are urgently needed [4,7].

Detailed experimental measurements are required to resolve boundary layer physics and provide valuable databases for model development. Hybrid particle image velocimetry (PIV) and particle tracking velocimetry (PTV) techniques have significantly advanced the level of detail in which engine boundary layer flows can be experimentally investigated. Previous investigations [5,8] utilized PTV to reveal several sub-millimeter vortices moving through the outer boundary layers in an ICE. Notable simulations of these experiments also demonstrated that the large-scale tumble flow introduces substantial pressure gradients in the outer boundary layer [9]. These unsteady flow features demonstrated significant deviations from common equilibrium model predictions, while significant improvement was achieved using non-equilibrium wall models [9,10].

Previous experiments and simulations have focused on non-reacting boundary layer flows, while combustion cases remain limited. Foster and Witze [11] used laser Doppler anemometry (LDA) to measure boundary layer thickness after combustion. Combustion yielded larger boundary layer thickness than motored operation, primarily due to an increase in fluid viscosity. Due to inherent limitations of LDA measurements, detailed analysis of the interaction between the flame and flow was not possible. The transient behavior of the bulk- and boundary layer flow in response to an advancing flame front must be well-understood to predict detailed flame propagation near engine surfaces [12].

The aim of this work is to study the transient near-wall flow development associated with an approaching flame front. Flow-field and flame distributions were measured simultaneously at kHz repetition rates using PTV and planar laser induced fluorescence (PLIF) of sulfur dioxide ($SO_2$). Measurements were performed near the piston surface of an optically accessible engine. Analyses focus on flame development for different velocity distributions existing before ignition. A strong interdependency between flow and flame development is shown which subsequently influences combustion. Near-wall velocities tangent to the piston surface are correlated with in-cylinder pressure to evaluate combustion performance. Unsteady flow phenomena associated with flame development are observed, which significantly affect boundary layer development.

## 2. Experimental

### 2.1. Engine

Measurements were conducted in a single-cylinder spark-ignition optical engine. The engine is equipped with a 4-valve pentroof cylinder head, centrally-mounted spark plug, and quartz-glass cylinder and flat piston. The engine was not operated with induced swirl. Further details of the engine and inlet flow characterizations are described in [13-15]. The engine operated at 800 rpm with port-fuel injection of isooctane. Additional operating conditions are shown in Table 1.

Table 1: Engine operating conditions.

| Bore / stroke | 86 mm / 86 mm |
|---|---|
| Geometrical compr. ratio | 8.7 |
| Engine speed | 800 rpm |
| Avg. intake pressure | 0.95 bar |
| Avg. intake temperature | 310.8±0.2 K |
| Fuel, equivalence ratio | Isooctane, 1.0 |
| Spark timing, dwell | -14.2°CA, 3.5 ms |
| IMEP, CoV | 5.3 bar, 1.9 % |

Figure 1 shows the average in-cylinder pressure trace. Ignition timing is highlighted and the fired pressure trace deviates from the motored trace around −7°CA (crank angle degrees are referenced to top-dead-center compression). The standard deviation indicates cyclic variations which lead to a variation in both maximum in-cylinder pressure ($P_{max}$) and the location of $P_{max}$.

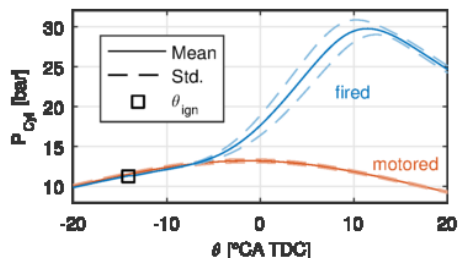

Fig. 1. Average in-cylinder pressure of fired and motored operation (200 cycle statistic for each operation).

*2.2. Optical setup*

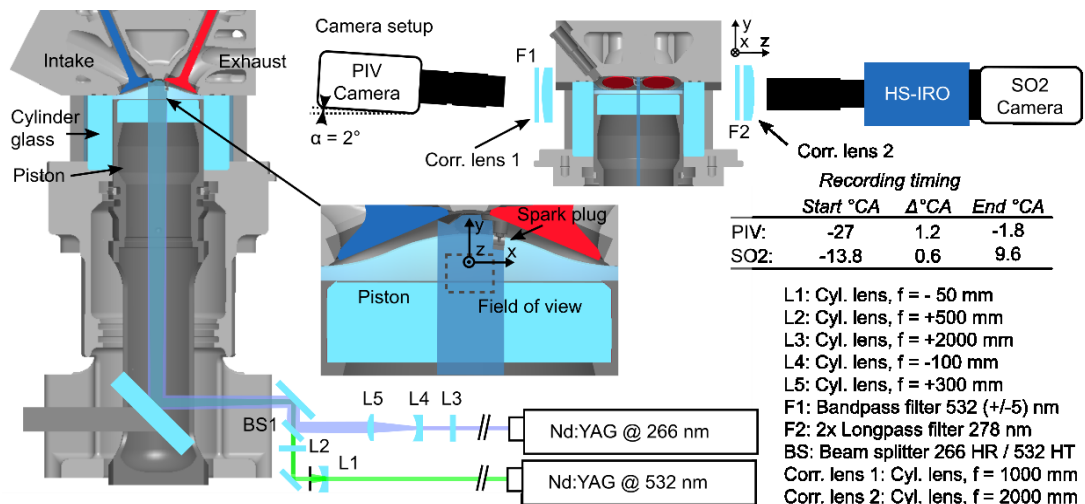

Fig. 2. Experimental setup.

Figure 2 shows the experimental setup. Tracer PLIF of $SO_2$ was used for flame imaging. $SO_2$ fluorescence shows a strong increase with temperature upon excitation with UV light [16]. This has already been successfully demonstrated for flame imaging in engines [17]. $SO_2$ was seeded into the intake air at a concentration of 1.1vol% $SO_2$. A frequency-quadrupled Nd:YAG laser (Edgewave, INNOSLAB) excited $SO_2$ with repetition rate of 8.2kHz (0.6°CA) and energy of 0.36mJ/pulse. Laser light was formed into a sheet (15mm width, 180μm FWHM thickness) and vertically guided into the engine.

Fluorescence signals were detected in the engine symmetry plane using an image intensifier (HS-IRO, LaVision) with 100ns exposure to suppress flame luminosity. A CMOS camera (Phantom v711, Vision Research) recorded the intensified signals. The intensifier was equipped with a 150mm lens (Halle, f# = 2.5) and 145mm extension rings providing a magnification of M=1.6 (80.6pxl/mm). A corrective lens was used to correct optical aberrations induced by the cylinder glass [18]. The optical resolution (101μm) was determined using a Siemens star target with a local median filter applied to target images. The PLIF system recorded a 12 x 9 mm$^2$ field-of-view (FOV) near the piston surface. The FOV is slightly offset from the cylinder axis towards the spark plug where the flame can be studied during its early development phase. The slight offset with the cylinder axis maintained sufficient optical correction with the large magnification.

The flow-field was measured in the central symmetry plane using a hybrid PIV/PTV technique. Silicone oil droplets (0.5μm diameter) were seeded into the intake air. The particle response time was calculated to be $t_p$=0.42μs for the in-cylinder conditions after ignition [19]. This is negligible compared to the laser pulse separation of dt=8μs. The particles were illuminated with a frequency doubled Nd:YAG laser (Edgewave, INNOSLAB). A laser sheet was formed (20mm width, 60μm FWHM thickness) overlapping the same path as the UV laser and guided into the engine. Mie scattering was detected using a CMOS camera (Phantom v711, LaVision) equipped with a 180mm macro lens (Sigma, f#=8) and 45mm extension rings. The camera was slightly tilted vertically to reduce vignetting at the piston. A corrective lens was used to reduce optical aberrations due to astigmatism. The magnification was M=1.9 (95pxl/mm).

All cameras and lasers were synchronized to the crank shaft with a timing unit (HSCv2, LaVision). For a given recording, five work/rest sequences (250 fired and 350 motored cycles) were performed and measurements were recorded for the last 100 fired cycles of this sequence. Flow-field images were recorded from −27°CA to −1.8°CA every 1.2°CA (4.1 kHz), while PLIF images were acquired from −13.8°CA to 9.6°CA every 0.6°CA (8.2 kHz). PTV measurements were also recorded for motored (non-fired) cycles acquired after each fired-cycle sequence. Flow-field images were acquired for a total of 60 motored cycles, when engine surfaces remained hot and exhaust temperatures showed a decrease of 7°C to the maximum fired temperature.

*2.3. Data processing*

Images of a spatially defined target (LaVision) within the engine were used to calibrate PLIF and PTV images using a 3$^{rd}$ order polynomial fit to match viewing planes of each system. PLIF images were processed in MATLAB. Background correction was applied using the signal of non-fired cycles. A sheet correction was performed when the flame filled the FOV. A 13x13pxl$^2$ (161μm) moving median filter was applied to the flame images. Images were then binarized with local, adaptive thresholds determined within a moving 50x50pxl$^2$ window. The local probability of burnt gas (referred to as "flame pdf") was calculated from binary images that identified the burnt gas at fixed CA.

Flow-field measurements were calculated using Davis 8.4.0 with a hybrid PIV/PTV algorithm [20]. Image preprocessing included a moving background subtraction and particle normalization (9 pxl window size). PIV vectors were calculated with decreasing window size (64 to 32 pxl), 75% overlap, and adaptive interrogation window shape. Vectors with correlation values below 0.1 were removed. PTV was calculated for a particle size range of 2-5pxl and a correlation window size of 8pxl. The same vector post-processing was applied as for the PIV steps. A denoising filter was applied using a polynomial fit of 2$^{nd}$ order.

Binarized flame images were used to mask the location of the flame in PTV images. For some images, out-of-plane flame locations caused significant beam steering causing particle locations to become defocused. These regions were identified using a local standard deviation filter. Vectors at positions where the local standard deviation was low (indicating a weak particle signal) were removed. The piston position was determined using phase averaged images from which the peak intensity location was determined [21].

Unstructured PTV vectors are mapped onto a regular mesh to present flow-field statistics at a resolution superior to PIV [22]. The average vector distance in the images was 110μm. Velocity vectors were spatially averaged onto a Cartesian mesh with Δx/Δy=0.5mm/0.05mm, providing more than one vector per cell and cycle on average.

## 3. Near-wall flame/flow imaging

Figure 3 presents simultaneous PLIF and PTV image sequences to describe the evolution of the flow-field and flame propagation as the flame approaches the piston surface. The spark plug is located at x/y = 6mm/2.4mm (i.e. beyond the upper right corner). Velocity vectors are colored by velocity magnitude, while the $SO_2$ signal is shown in gray-scale. Two high-speed image sequences are shown to describe distinct trends observed in the measurements. The top row shows a cycle with high velocities (6-11 m/s) above the piston surface at −15°CA (before ignition). The flow is directed towards the –x direction (right-to-left) and illustrates a high velocity 'sweep' and 'eject' phenomena known as 'bursting' [23]. These flows are known for significant mass and momentum transfer between inner and outer layers of the boundary layer. As the flame enters the FOV, the high-velocity flow persists and is directed parallel to the piston surface. As the flame progresses, velocity increases ahead of the flame, most notably due to gas expansion behind the flame that accelerates the unburnt gas.

The bottom image sequence shows a cycle with significantly lower velocity magnitude (2-4 m/s) before ignition. In comparison, this flow does not exhibit a strong right-to-left 'sweeping flow'. Instead the flow is directed more towards the +y direction. The flame progresses much slower through the FOV and flow direction parallel to the piston is eventually formed. As a result of this slower flame propagation, there is only a mild increase of the velocity magnitude downstream of the flame.

Two interesting observations are derived from these imaging sequences: (1) a mutual relationship between near-wall flow and flame progression exists; flame progression in the FOV is faster with higher initial velocity and velocity increases further as the flame advances. This relationship may affect combustion performance. (2) A flame approaching the wall produces an apparent flow acceleration ahead of the flame. This unsteady nature of the bulk- and near-wall flow will no-doubt influence mass and momentum transfer in boundary layers. The following analyses focus on these aspects.

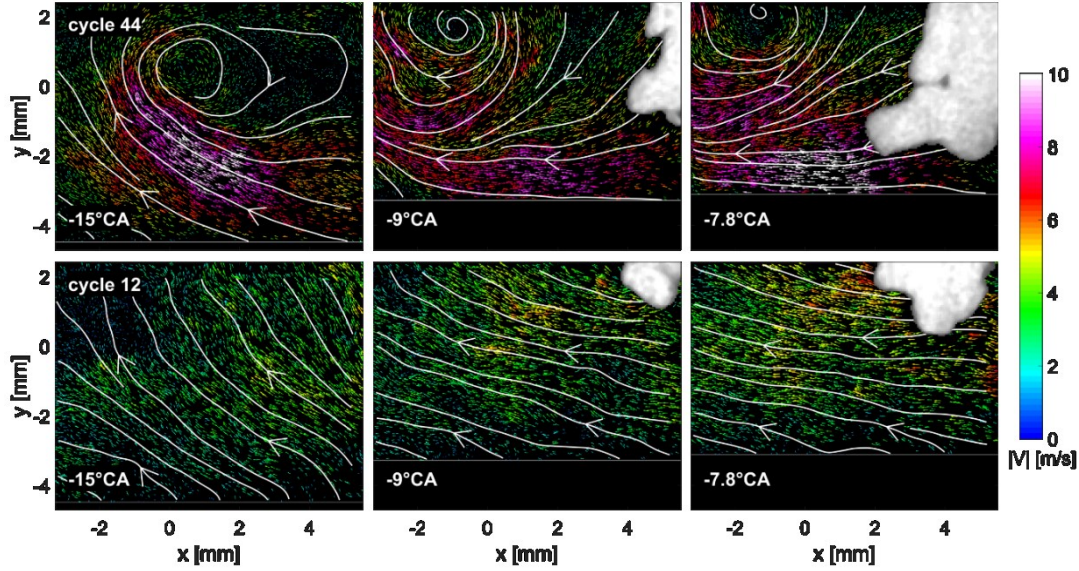

Fig. 3. Instantaneous flame/flow imaging showing a sequence with strong flow (top) and weak flow (bottom). Vectors demonstrate the high spatial resolution and superimposed streamlines indicate the flow direction.

*3.1. Flow/combustion correlation*

A correlation analysis was conducted to understand the relationship between near-wall flow and combustion. Figure 4 shows a 2D correlation map between horizontal velocity ($V_x$) spatially averaged over a 0.5x0.5mm$^2$ window and maximum cylinder pressure ($P_{MAX}$). The correlation coefficient was calculated using:

$$\sum_{i=1}^{n}[(V_{x,i} - \overline{V_x})(P_{max,i} - \overline{P_{max}})]/\sigma_{V_x}\sigma_{P_{max}} \qquad (1)$$

Where *i* indicates the cycle, the overbar indicates the average value of the full sample and $\sigma$ is the standard deviation. This analysis focuses on flow images at −9°CA when the flame enters the FOV and is based on 200 engine cycles.

$V_X$ was chosen in this analysis because this was the dominant flow direction as the flame propagated through the FOV. Negative $V_X$ values indicate velocity in the –x direction. $P_{MAX}$ was chosen because it showed high correlation values and is a good overall combustion indicator. For the operating conditions employed, a higher $P_{MAX}$ indicated an earlier $P_{MAX}$ location.

The insert in Fig. 4 shows the values from within the blue rectangle. It indicates a strong negative correlation between $P_{MAX}$ and $V_X$ (i.e. stronger flow in the –x direction ($-V_X$) correlate to higher $P_{MAX}$). This equally indicates that a stronger $-V_X$ gives a faster propagating flame front, yielding quicker combustion. The correlation map shows that at $-9°CA$, $V_X$ values showing the strongest correlation are located directly above the piston surface. At later °CA (not shown), high correlation values are shifted from right-to-left similar to the flame growth direction. This analysis demonstrates a strong relationship between the near-wall flow and subsequent combustion.

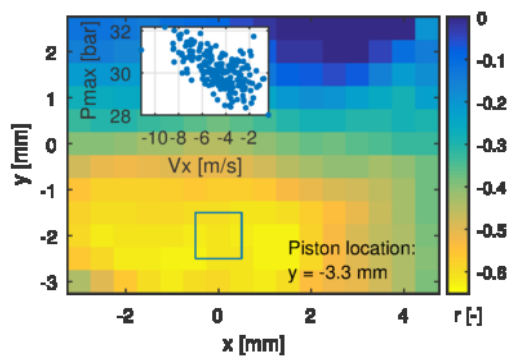

Fig 4. 2D correlation map between $V_X$ and $P_{MAX}$ at $-9°CA$ (200 cycle statistic). Insert displays average $V_X$ of the blue rectangle.

## 4. Conditional averaging analysis

This section presents conditional average findings based on cycles with high and low $V_X$ above the piston surface (Fig. 4, blue rectangle). Figure 5 shows conditionally averaged flow-fields and flame statistics for 4 data sets.

- A: 200 fired cycles
- $A_W$: 100 cycles from A with weak flow ($V_X > -4.3$m/s)
- $A_S$: 100 cycles from A with strong flow ($V_X < -4.3$m/s)
- $A_M$: 60 motored cycles

Accordingly, cycles are classified as strong (weak) if the flow magnitude within the blue rectangle of Fig. 4 is higher (lower) than 4.3m/s in negative x-direction. Flow statistics are only based on the unburnt gas. Velocity data is resampled on the aforementioned structured mesh (Δx/Δy=0.5mm/0.05mm), giving approx. 400 vectors per cell in the bulk flow and 100 vectors per cell closest to the wall. Data in the upper right-hand corner consists of 50-100 vectors per cell, because these cells often contain burnt gas. Flame statistics, determined from binarized PLIF images, are shown by contour lines indicating burnt gas locations with 5, 25 and 50% probability.

Differences in the flow-field between $A_W$ and $A_S$ are already present at $-15^oCA$ (before ignition) even though subsets were sampled by $V_X$ at $-9^oCA$. Although differences in flow pattern between $A_W$ and $A_S$ are small, velocity magnitude is two times larger for $A_S$. $A_M$ shows a similar flow pattern to fired subsets, and velocity magnitude is similar to the full data set A.

As the flame enters the FOV, the flow direction becomes more parallel to the piston surface and velocity magnitude increased for all fired subsets. Compelling differences in flame propagation and flow-field are evident between the $A_W$ and $A_S$ subsets. Flame propagation, indicated by PDF contours, is much faster for the $A_S$ subset. At $-7.8^oCA$, a flame can already impinge on the piston surface, while this is not the case for $A_W$ cycles. While both subsets show flow acceleration ahead of the flame, velocity magnitude remains higher for $A_S$ than $A_W$.

In agreement with Fig. 3, conditional analysis further reveals a two-way cause-and-effect relationship between the flame and flow-field. Faster flame development can be associated with higher initially velocity magnitudes. It is anticipated that flow velocity is not only faster near the wall, but also at other chamber locations, which can be a driving force for faster flame propagation. This phenomena can be seen in previous flame/flow studies within this engine [24-26]. At the same time, flame propagation in the FOV also accelerates unburnt gas downstream the flame, generating higher velocities. For the $A_W$ subset, the unburnt gas flow acceleration above the piston is slower due to retarded flame propagation. Although, the unburnt flow eventually accelerates, flow velocities remain about two times slower for $A_W$ than for $A_S$.

It is compelling that the motored subset ($A_M$) exhibits a decreasing flow velocity with increasing CA. Additionally, the motored flow does not exhibit such a strong parallel flow alignment with the piston surface. It is evident that higher velocities parallel to the piston surface are a direct result from flame expansion and that this will have a strong impact on boundary layer development and the requirements to properly predict it.

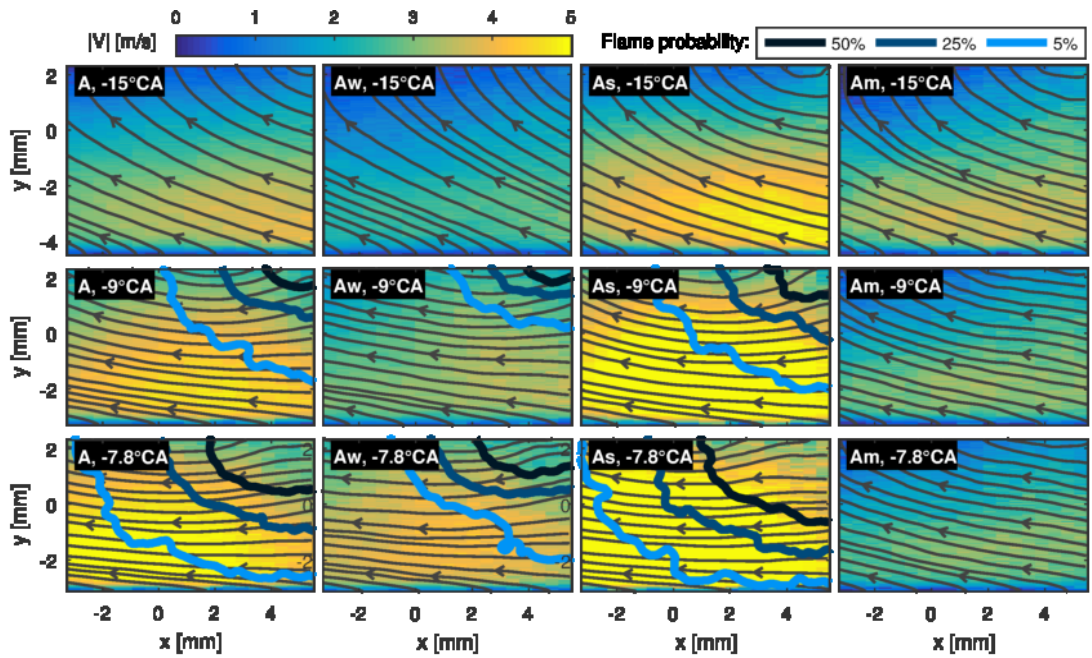

Fig. 5. Average flow-field and flame pdfs. Bottom of image indicates piston location.

*4.1. Velocity profiles normal to piston*

Figure 6 presents $V_X$ profiles normal to the piston surface ($y'$) to study trends in boundary layer development. Velocities are reported from the 0.5mm/0.05mm (x/y) mesh spacing. Velocities are spatially averaged between x=−0.25 to 0.25mm for each 0.05mm increment normal to the wall. Figure 6 reports ensemble-average $V_X$ velocities of *unburnt* gas (every 4$^{th}$ data point highlighted) with the first data point at $y'$=0.025mm from the piston. Bars, indicating the 95% confidence interval of the mean ($\Delta V_{X,95\%}$), are largest at $y'$=0.025mm due to reduced number of vectors and moderate variations in $V_X$. The location of $\delta_{75}$=0.75$V_{X,MAX}$ is indicated in Fig. 6 to discuss qualitative trends of boundary layer development between different subsets and recorded CA. The value $\delta_{99}$, often identifying the boundary layer thickness for steady flows [27], is not reported because of the unsteady nature of the bulk flow. In-cylinder pressures are reported in Fig. 6. Standard deviations were below 0.1bar at each CA presented (see also Fig. 1) and differences between the datasets were below 0.1bar.

The $\delta_{75}$ values reported in Fig. 6 include horizontal bars that indicate the range of $\delta_{75}$ values associated with the confidence interval of the mean $V_X$ values (termed $\Delta V_{X,95\%}$). These horizontal bars represent the $\delta_{75}$ confidence interval. In some instances these bars are large and require interpretation. The reader is reminded that the flow is turbulent and the flow structure can exhibit large variation from cycle-to-cycle. Furthermore, the bulk velocity is not spatially uniform (e.g. vortex shown in Fig. 3). These aspects yield considerable $\Delta V_{X,95\%}$ values (i.e. vertical bars). The length of the horizontal bars depends both on the gradient of the mean $V_X$ profiles and the magnitude of $\Delta V_{X,95\%}$. When the $\delta_{75}$ position is located at low mean $V_X$ gradients and large $\Delta V_{X,95\%}$ magnitudes, the horizontal bars become notably long especially towards higher $y'$ values. This is particularly true for the motored dataset at -9.0 and -7.8°CA where $V_X$ values are amongst the lowest and $\Delta V_{X,95\%}$ values are the largest.

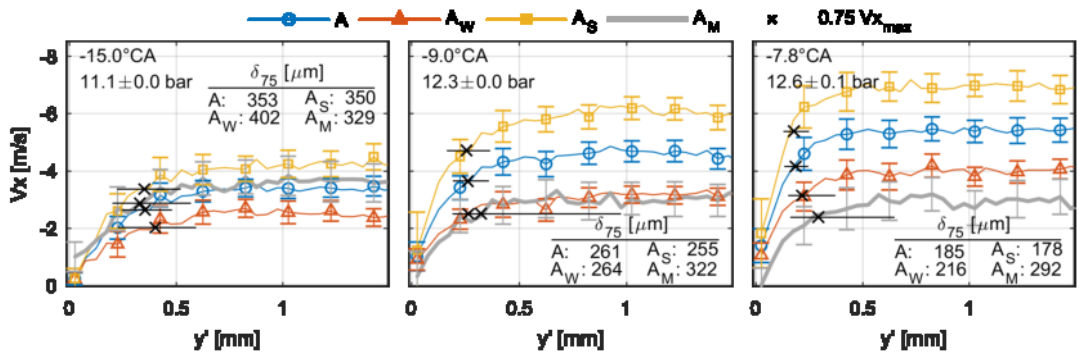

Fig. 6. Ensemble-average $V_X$ normal to the piston surface. Bars denote 95% confidence interval of the mean. Symbols indicate every 4th data point, while lines interpolate between measurement points. $y'$ denotes the normal direction relative to the piston surface.

Velocity profiles reveal consistent findings to those presented in Fig. 5, but additionally provide a qualitative comparison of boundary layer development. At −15°CA, average $V_X$ profiles are similar for subsets A, $A_S$, and $A_M$. Consequently the $\delta_{75}$ location is also consistent between these subsets for the reported $\delta_{75}$ confidence intervals.

At −9°CA, flame propagation has significantly increased velocities for $A_S$ and A subsets, while slower flame propagation produces a mild $V_X$ increase for $A_W$. Similar velocity profiles for $A_W$ and $A_M$ indicate that $A_W$ has velocities resembling those of the mean non-reacting flow. At −9°CA, mean $\delta_{75}$ values have decreased for all subsets. The increase in pressure from −15°CA to −9°CA will contribute to this decreased $\delta_{75}$ value. However, all fired subsets also exhibit an increase in velocity for which $\delta_{75}$ values are similar. At −7.8°CA, flame propagation increases $V_X$ for all fired datasets within the short time duration, while $V_X$ decreases for $A_M$. $\delta_{75}$ values are smallest for the fired subsets (see numerical values within inserts of Fig. 6). The pressure change between −9°CA and −7.8°CA is small (~0.3bar) which may suggest that the decrease in $\delta_{75}$ may be attributed to the velocity increase. However, $A_M$ also exhibits a decrease of $\delta_{75}$ at −7.8°CA (albeit not as drastic), but a decrease in velocity. This observation is inconsistent with the fired subsets. Larger statistics from more cycles (especially motored limited to 60 cycles) and at additional CA would help evaluate these trends further.

Flow acceleration, attributed to flame expansion, rapidly increases velocity gradients and the boundary layer development becomes highly transient. Table 2 reports $Vx_{75}/\delta_{75}$ to emphasize changes in velocity gradients with time. While this parameter under-predicts maximum gradients, it can provide a meaningful value for engine simulations and resolution requirements at walls. Gradient values increase with increasing °CA for all fired subsets with the largest gradients reported for $A_S$. Most notable is the strong deviation from the motored flow, which decelerates in time. This emphasizes the need to properly model the additional transient aspects of boundary layer development for reacting flows within numerical simulations.

Table 2: Velocity gradients $Vx_{75}/\delta_{75}$ normal to the piston.

| | $Vx_{75}/\delta_{75} \times 10^{-3}$ [1/s] | | | |
|---|---|---|---|---|
| Θ [°CA] | A | $A_W$ | $A_S$ | $A_M$ |
| -15 | -7.4 | -5.0 | -9.6 | -8.4 |
| -9 | -14 | -9.5 | -18 | -7.8 |
| -7.8 | -22 | -15 | -30 | -8.0 |

## 5. Conclusions

Simultaneous measurements of PTV and SO$_2$ PLIF are utilized to study the interaction between flow and flame development near the piston surface in an optically accessible SI engine. The engine operated at 800 rpm with premixed stoichiometric isooctane-air mixtures. Measurements were acquired at 4.1 kHz (PTV) and 8.2 kHz (PLIF) repetition rates to resolve the transient flow behavior in response to an approaching flame front. Additional PTV measurements were taken without combustion to compare near-wall flow development without the presence of a propagating flame.

High-speed measurements revealed a strong interdependency between near-wall flow and flame development. A flame front progresses faster for higher initial flow velocities near the wall. This faster flame propagation accelerates the unburnt gas ahead of the flame, further increasing flow velocities parallel to the piston surface. As a flame enters the FOV, large $V_X$ velocity magnitudes are correlated with a faster propagating flame front, which yields faster combustion and higher $P_{MAX}$. Largest correlation values are shown for flow velocities directly above the piston surface indicating a compelling relationship between near-wall flow velocity and subsequent combustion.

A conditional analysis was performed to study flame/flow interdependencies at the piston surface for cycles with 'weak' and 'strong' flow parallel to the piston surface. Faster flame propagation, associated with higher flow velocities before ignition, demonstrated a stronger flow acceleration ahead of the flame. Cycles with low velocities demonstrated slower flame propagation, which yielded a weaker flow acceleration. Flow acceleration associated with an advancing flame front is a transient feature that significantly affected boundary layer development. $\delta_{75}$ qualitatively compared boundary layer development for different subsets. Flow acceleration, accredited to flame expansion, rapidly increased velocity gradients at the wall, and $\delta_{75}$ revealed a decreasing trend with increasing flow velocity for fired subsets. Most intriguing was a strong deviation of the boundary layer flow between fired and motored subsets, the latter of which did not exhibit strong transient flow behavior.

The interaction of flow and flame close to the wall shows that sufficiently resolving the boundary layer is of great importance on the path towards predictive simulations. The results of this study can give a good comparison to simulations but also show that more studies (numerical and experimental) are needed to assess the complex inter-dependencies within ICE combustion.

Further measurements will include wall temperature measurements that will help calculate wall shear stress and transient heat flux measurements. Such measurements are intended to provide a better understanding of boundary layer development for reacting flows in IC engines.

**Acknowledgements**

We kindly acknowledge financial support through SFB/Transregio 150 of the Deutsche Forschungsgemeinschaft. A. Dreizler is grateful for generous support by the Gottfried Wilhelm Leibniz program. B. Peterson gratefully acknowledges financial support from the European Research Council (ERC grant #759546).

**References**

1. G. Borman, K. Nishiwaki, *Progress in Energy and Combustion Science* 13 (1) (1987) 1–46.
2. J.B. Heywood, *Internal Combustion Engine Fundamentals*, McGraw-Hill, New York, U.S.A, 1988.
3. Z. Han, R.D. Reitz, *International Journal of Heat and Mass Transfer* 40 (3) (1997) 613–625.
4. C.J. Rutland, *International Journal of Engine Research* 12 (5) (2011) 421–451.
5. C. Jainski, L. Lu, A. Dreizler, V. Sick, *International Journal of Engine Research* 14 (3) (2012) 247–259.
6. M. Schmitt, C.E. Frouzakis, Y.M. Wright, A.G. Tomboulides, K. Boulouchos, *International Journal of Heat and Mass Transfer* 91 (2015) 948–960.
7. J. Borée, P.C. Miles, in: C. David, D.E. Foster, T. Kobayashi, N. Vaughan (Eds.), *Encyclopedia of Automotive Engineering*, John Wiley & Sons, Ltd, Chichester, U.K., 2014.
8. A.Y. Alharbi, V. Sick, *Exp Fluids* 49 (4) (2010) 949–959.
9. P.C. Ma, T. Ewan, C. Jainski, L. Lu, A. Dreizler, V. Sick, M. Ihme, *Flow Turbulence Combust* 98 (1) (2017) 283–309.
10. P.C. Ma, M. Greene, V. Sick, M. Ihme, *International Journal of Engine Research* 18 (1-2) (2017) 15–25.
11. D.E. Foster, P.O. Witze, *SAE Technical Paper 872105* (1987).
12. O. Laget, L. Muller, K. Truffin, J. Kashdan, R. Kumar, J. Sotton, B. Boust, M. Bellenoue, *SAE Technical Paper 2013-01-1121* (2013).
13. D. Freudenhammer, B. Peterson, C.-P. Ding, B. Boehm, S. Grundmann, *SAE Int. J. Engines* 8 (4) (2015) 1826–1836.
14. E. Baum, B. Peterson, B. Böhm, A. Dreizler, *Flow Turbulence Combust* 92 (1-2) (2014) 269–297.
15. F. Zentgraf, E. Baum, B. Böhm, A. Dreizler, B. Peterson, *Phys. Fluids* 28 (2016) 045108.

16. K. Fukui, T. Fujikawa, M. Tohyama, Y. Hattori, K. Akihama, *SAE Int. J. Engines* 6 (1) (2013) 289–299.

17. R. Honza, C.-P. Ding, A. Dreizler, B. Böhm, *Appl. Phys. B* 123:246 (9) (2017).

18. M. Reeves, *Particle Image Velocimetry Applied to Internal Combustion Engine In-Cylinder Flows*, PhD thesis, Loughborough University, Loughborough, U.K., 1995.

19. C. Tropea, A.L. Yarin, J.F. Foss, *Springer Handbook of Experimental Fluid Mechanics*, Springer, Berlin, Germany, 2007.

20. M. Stanislas, K. Okamoto, C.J. Kähler, J. Westerweel, *Exp Fluids* 39 (2) (2005) 170–191.

21. J.R. MacDonald, C.M. Fajardo, M. Greene, D. Reuss, V. Sick, *SAE Technical Paper 2017-01-0613* (2017).

22. C.J. Kähler, S. Scharnowski, C. Cierpka, *Exp Fluids* 52 (6) (2012) 1641–1656.

23. S.K. Robinson, *Annu. Rev. Fluid Mech.* 23 (1) (1991) 601–639.

24. B. Peterson, E. Baum, B. Böhm, A. Dreizler, *Proc. Combust. Inst.* 35 (2015) 3829-3837.

25. B. Peterson, E. Baum, A. Dreizler, B. Böhm, *Combust. Flame* (submitted).

26. C. He, G. Kuenne, E. Yildar, J. van Oijen, F. di Mare, A. Sadiki, C.-P. Ding, E. Baum, B. Peterson, B. Böhm, J. Janicka, *Combustion Theory and Modelling* 21:6 (2017) 1080-1113.

27. H. Schlichting, K. Gersten, *Boundary-Layer Theory*, Springer, Berlin, Germany, 2017.